# Quantitative Relationship between Population Mobility and COVID-19 Growth Rate based on 14 Countries

Benjamin Seibold[1], Zivjena Vucetic[2], Slobodan Vucetic[3]

[1] Department of Mathematics, Temple University, PA, USA
[2] Clinical Genomics, NJ, USA
[3] Department of Computer and Information Sciences, Temple University, PA, USA

**Abstract**
This study develops a framework for quantification of the impact of changes in population mobility due to social distancing on the COVID-19 infection growth rate. Using the Susceptible-Infected-Recovered (SIR) epidemiological model we establish that under some mild assumptions the growth rate of COVID-19 deaths is a time-delayed approximation of the growth rate of COVID-19 infections. We then hypothesize that the growth rate of COVID-19 infections is a function of population mobility, which leads to a statistical model that predicts the growth rate of COVID-19 deaths as a delayed function of population mobility. The parameters of the statistical model directly reveal the growth rate of infections, the mobility-dependent transmission rate, the mobility-independent recovery rate, and the critical mobility, below which COVID-19 growth rate becomes negative. We fitted the proposed statistical model on publicly available data from 14 countries where daily death counts exceeded 100 for more than 3 days as of May 6$^{th}$, 2020. The publicly available Google Mobility Index (GMI) was used as a measure of population mobility at the country level. Our results show that the growth rate of COVID-19 deaths can be accurately estimated 20 days ahead as a quadratic function of the transit category of GMI (adjusted $R^2$ = 0.784). The estimated 95% confidence interval for the critical mobility is in the range between 36.1% and 47.6% of the pre-COVID-19 mobility. This result indicates that a significant reduction in population mobility is needed to reverse the growth of COVID-19 epidemic. Moreover, the quantitative relationship established herein suggests that a readily available, population-level metric such as GMI can be a useful indicator of the course of COVID-19 epidemic.

**Introduction**

The ongoing SARS-CoV-2 pandemic, first detected at the end of December 2019 in Wuhan, China, has caused substantial morbidity and mortality globally. In the five months since the beginning of the outbreak, 188 countries/regions had reported the COVID-19 disease and there have been around 350,000 confirmed deaths and over 5.6 million confirmed cases globally [1, 2].

In the absence of a vaccine and effective medical treatment, unprecedented national and community non-pharmaceutical, social distancing interventions (NPIs), including travel restrictions and border closures, limitations on gatherings and public events, and school, workplace and business closures, were introduced across the globe in addition to recommendations about personal hygiene and behavior (washing hands, covering cough, keeping 6-feet distance, wearing masks, self-isolation)[3]. The goal of NPIs is to reduce the transmission of the SARS-CoV-2 virus and thus slow the exponential growth of the epidemic [4]. Delaying the onset and reducing the magnitude of the infection peak helps to better utilize medical resources, to avoid the overburden of the health care system and to effectively "buy time" until a cure and/or vaccine is available. While reducing the impact of COVID-19 on population morbidity and mortality is the desired health outcome, the economic and social consequences of social distancing interventions are devastating. Therefore, it is critical to understand if social distancing interventions and policies introduced by governments and public health authorities are effective in achieving desirable health outcomes. Even more, the ideal scenario would be to establish a distinct quantitative relationship between a readily available metric of the population's response to policies in place and the growth rate of the epidemic.

One such metric is population mobility, which can be assessed and quantified either on an individual or a population level and compared before and after the implementation of various NPIs. The first studies about the impact of social distancing and mobility on COVID-19 epidemic came from China. A study by Kraemer et al. [5] used mobility data and fine-grained epidemiological data at the level of individual people to explain the spread and growth rate in early stages of the COVID-19 epidemic in China. The study found that variation in the COVID-19 growth rates outside of Wuhan could be almost entirely explained by population movement from Wuhan. The study also explained how drastic social distancing and the control of movement measures across the country resulted in negative growth rates. A related work that used fine-grained mobile phone data showed that movement of undocumented cases contributed to over 80% infections during the early stages of the epidemic in China [6]. Despite the useful insights, it is difficult to replicate these studies in other countries due to obstacles in obtaining fine-grained, individual-level linked mobility and case data.

Several publications report on the relationship between implementation of social distancing measures and population mobility. Klein et al. [7] used Cuebiq company aggregated and de-identified data collected from 17 million mobile devices between January, 1 and March 25, 2020, to conclude that within a week of implementation of social distancing, the major U.S. metropolitan areas experienced on average 50% reduction in typical commutes to/from work. A similar study in Italy [8] showed that during the early phase of the outbreak when only mild measures were in place (e.g., school closures) the traffic to/from the most affected provinces declined by about 50%. Recently, the public release of population mobility data by Google (as Google Mobility Index, GMI) [9] and Apple (as Mobility Trends Report) made it possible to understand how population mobility changed across the globe in response to social distancing interventions. While it is clear that most countries accomplished a significant reduction in population mobility, elucidating the precise relationship between population mobility and COVID-19 growth is still an open question.

There are in general two types of studies attempting to model and predict behavior of COVID-19 epidemic. The first type of studies relies on mathematical models that realistically reproduce disease dynamics. For example, Chinazzi et al. [10] used a stochastic disease transmission model to show that strict travel restrictions delay the disease spread by only a few days. Linka et al. [11] used a network simulation model to understand the impact of travel restrictions on COVID-19 spread in Europe; and Tuite et al. [12] used a mathematical model to study the epidemic spread in Canada. The mathematical models rely on first principles that specify fundamental dynamic of an epidemic typically require multiple parameters, some of which are selected based on the known information about the COVID-19 and others are fitted using historical data. The resulting dynamics for all but the simplest models cannot be described in a closed form, which reduces their interpretability. Despite calibrating their parameters to data and constraining their values based on domain knowledge, they can still overfit and produce incorrect behaviors if the number of parameters is large compared to the size of available data. Another type of studies [13, 14] develops statistical prediction models that are simple to describe but are not rooted in first principles. A representative of such an approach is a widely publicized Institute for Health Metrics and Evaluation (IHME) model [15] which assumes that the logarithm of positive cases follows a quadratic function with the global maximum and thus optimistically predicts a quick decline of an epidemic. The UTexas model [16] extends the IHME model by allowing the quadratic function to be governed by the population mobility. It is notable as one of the few studies that directly link population mobility and number of cases. A related approach [17] provides a more elaborate model with more realistic assumptions but a larger number of parameters that are fitted using Bayesian approaches. A common weakness of statistical models is that they are not derived from first principles such that, while they can result in accurate short-term forecasting, they underperform in reproducing the natural behavior of an epidemic and are unreliable for longer-term forecasting. Our objective is to combine a mathematical model of epidemic dynamics and a statistical prediction model, to elucidate a quantitative relationship between COVID-19 infection growth rate and mobility as a proxy of social distancing.

**Framework for Modeling of COVID-19 Growth Rate**

As illustrated in Figure 1 our modeling framework is based on 3 interconnected models — SIR model, Social Distancing Model and Statistical Model (see Materials and Methods for details). The *augmented SIR (Susceptible-Infected-Recovered) epidemiological model* establishes the underlying dynamics of an epidemic where individuals can be classified by their epidemiological status as susceptible to the infection (S), infected and therefore infectious (I), and recovered and no longer infectious (R). The central quantity is the infection growth rate $\lambda_I(t)$, which is the relative rate of increase in total number of infected individuals $I(t)$ at time $t$ that can be calculated as $\lambda_I(t) = (I(t + 1) - I(t))/I(t)$. The infection growth rate depends on the transmission rate $\beta(t)$ (how quickly an infected person infects somebody else), the recovery rate $\gamma$ (how quickly an infected person recovers), and the susceptible ratio $\mu$ (what fraction of population is susceptible to becoming infected). In our framework, we assume we are observing an initial stage of a new epidemic, so the susceptible ratio can be assumed to be close to 1. Then, the infection growth rate is approximated as $\lambda_I(t) = \beta(t) - \gamma$ (see Materials and Methods). Under some additional assumptions, the growth rate $\lambda_j(t)$ in newly infected individuals $j(t)$ equals the growth rate $\lambda_I(t)$ in total infections, $\lambda_I(t) = \lambda_j(t)$. By augmenting the SIR model, it can be shown that the growth rate $\lambda_d(t + \Delta)$ in new fatalities $d(t + \Delta)$ at time $t + \Delta$ can be approximated by the growth rate $\lambda_j(t)$ in new infections at time $t$, $\lambda_d(t + \Delta) = \lambda_j(t)$, where $\Delta$ is the average delay (measured in number of days) between the time of infection and the time of death, if it occurs. Thus, it is also approximately correct that $\lambda_d(t + \Delta) = \lambda_I(t)$.

The *social distancing model* establishes that the transmission rate $\beta(t)$ depends on the behavior $b(t)$ of the impacted population, which is not easily observable. Population mobility $m(t)$ is an observable, standardized, and publicly available quantity that depends on the population behavior $b(t)$. The value $m(t) = 1$ corresponds to the baseline mobility immediately prior to the COVID-19 outbreak in that country, $m(t) = 0$ corresponds to the theoretical scenario when the population stops moving, and any number between those two extremes is a fraction of the baseline mobility. Because population mobility information is publicly available for multiple countries and regions (e.g. with Google Mobility Index) we use it as proxy for the behavior $b(t)$. We hypothesize that $m(t)$ can be used as a predictor of $\beta(t)$.

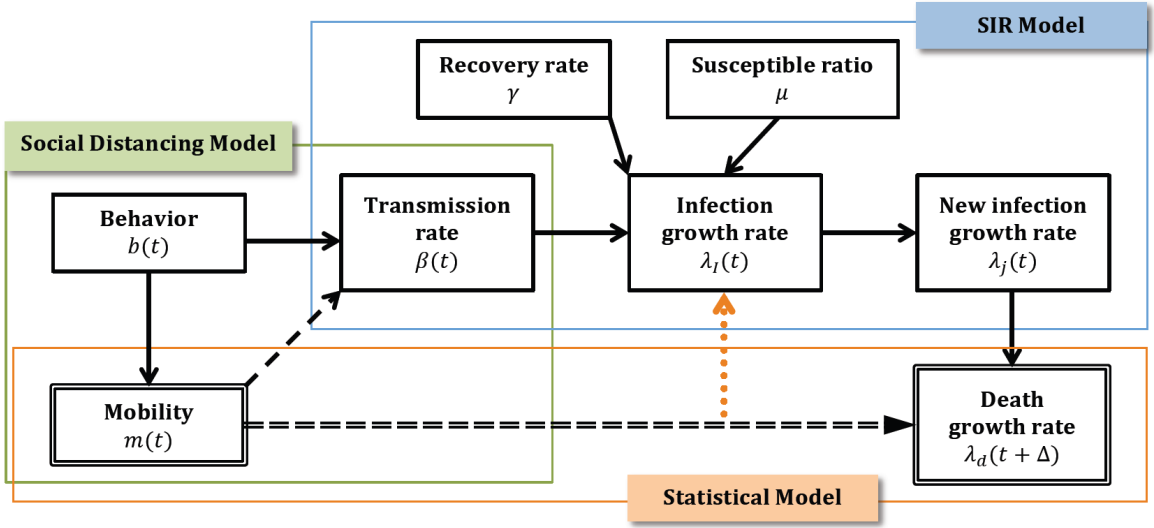

**Figure 1. Framework for modeling of COVID-19 growth rate.**

In Figure 1, the quantities in the boxes surrounded by single-line rectangles are either not readily observable (such as $b(t), \beta(t), \gamma$) or are unreliable (such as $\lambda_I(t), \lambda_j(t)$) and can be treated as hidden variables. The quantities in the boxes surrounded by double-line rectangles ($m(t)$ and $\lambda_d(t)$) are readily available and sufficiently reliable and can be treated as observed variables. The full-line arrows are causal relationships. The single-dashed line is an indirect causal relationship. By combining the social distancing model and the SIR model, the *statistical model* establishes a link (double-dashed line arrow) between the two observable variables: mobility $m(t)$ at time $t$ and death growth rate $\lambda_d(t + \Delta)$ at time $t + \Delta$ as $\lambda_d(t + \Delta) = f(m(t))$, where the function $f$ is fitted from the observed data about mobility and daily death counts using a regression algorithm. Since the augmented SIR model established that $\lambda_d(t + \Delta) = \lambda_I(t)$, $f(m(t))$ can be used to predict the current infection growth rate $\lambda_I(t)$ (orange dotted arrow). In addition, we can also directly estimate $\gamma$ and $\beta(t)$ as the intercept and the remaining terms of $f(m(t))$, respectively.

Our framework assumes that during the period covered by our data (from mid-February to May 9th, 2020) the 14 countries we analyzed were in the initial stages of COVID-19 epidemic, with large majority of population still being susceptible. Based on several population-based surveys that test for the presence of SARS-COV2 specific antibodies, the crude estimates of the proportion of individuals in the general population who developed antibodies and therefore have been infected is below 10% [18, 19]. Important exceptions are COVID-19 geographical hotspots like New York City where some reported prevalence of up to 24.7% or in very selected, small clusters of high-exposure risk populations like health-care workers and teachers, where estimates indicate that up to 50% or more individuals may have already been infected in some localities [18, 19].

The relationship $\lambda_d(t) = \lambda_j(t) = \lambda_I(t)$ is valid under the assumption that population mobility $m(t)$ changes slowly on a time scale of days. If $m(t)$ changes rapidly, as was the case during the first days of social distancing in most countries, it is difficult to write the relationship in a closed form. Due to this observation, before fitting the regression function $f$ we remove data points corresponding to days when population mobility changes rapidly.

The proposed framework relies on data about COVID-19-related deaths to infer the relationship between population mobility and COVID-19 growth rate. We note that using data about infection cases rather than deaths would result in a simpler estimation problem because it would obviate the need to use a delay $\Delta$ in the modeling. However, for several reasons, the publicly available numbers of confirmed COVID-19 cases provide a biased and unreliable view of the epidemic spread, due to the inability to accurately diagnose/identify all patients with the disease [20]. First, the majority of COVID-19 cases are only mildly symptomatic or even asymptomatic so they do not necessarily seek medical care; yet, these undocumented cases substantially contribute to the spread of the infection [6]. Second, laboratory testing for confirmation of the presence of the virus is the accepted method for verifying disease presence; however, it suffers from numerous sources of variability that are typically not reported in public data and are hard to control for. For example, testing capacity is a rate-limiting step for identifying new cases and it widely differs between countries. Similarly, different clinical criteria exist across countries that trigger lab testing so accessibility to testing varies across and within countries and across time. Further, numerous testing methods are used globally, all with different or unknown analytical accuracies for SARS-COV2 detection. Different test processing times (from swab to result) and delays in reporting of testing results (based on the day the test was administered or the day when the test results were received) add to the inconsistency and variability in reported number of daily new cases. Taken together, the inability to consistently determine the number of COVID-19 cases could not only underestimate true disease prevalence but also makes it difficult to compare across different countries and stages of the epidemic.

In contrast, although not perfect, the number of COVID-19-related deaths and death growth rate derived from it present a more reliable statistic for tracking the impact of COVID-19 across regions and times. Death reporting is generally standardized, and countries follow the "cause of death" classifications from the WHO's International Classification of Diseases guidelines [2, 21]. However, countries also have their own guidelines on how and when COVID-19 deaths should be recorded and the definitions of what is considered a "COVID-19 death" may have changed during the outbreak within a particular country [22, 23]. For example, as of April 14, 2020 US CDC expanded the definition of COVID-19 deaths to include both confirmed cases and "probable cases" [23], adding clinical and epidemiologic evidence as being independently sufficient instead of COVID-19 laboratory confirmation only. Several countries, like UK, Italy, Belgium and France [22, 24] were initially recording only in-hospital deaths but as the epidemic progressed they included out-of hospital deaths (i.e., nursing homes), too. Similar to case reporting, there are also delays and seasonality effects in death reporting, especially over the weekends and public holidays, where underreporting occurs during weekends and holidays followed by overreporting in the days after [25]. Additionally, for some countries the daily death count represents recorded death count on a particular day, not the actual number of deaths that occurred on that day. Our data preprocessing procedure is (see Materials and Methods) specifically developed to alleviate some of the issues related to death reporting.

## Results

### Data Description

We downloaded daily counts of COVID-19 deaths for 14 countries collected up to May 9, 2020 (see Table S1 for a list of country names and important events and statistics) from Our World in Data [2]. We downloaded population mobility data at the country level from Google's COVID-19 Community Mobility Reports (https://www.google.com/covid19/mobility/) [26], that are referred to as the Google Mobility Indices (GMIs). The data was released on May 9, 2020 and included mobility data from February 15, 2020 to May 3, 2020. For this study, we selected 14 countries if both population mobility and death counts data were available, and the death counts exceeded 100 for at least 3 days up to May 9, 2020.

As seen in Figure S1.a, there are 6 categories of GMI, measuring population mobility with respect to retail and recreational venues ("retail"), groceries and pharmacies ("grocery"), parks ("parks"), transit centers ("transit"), workplaces ("work"), and places of residence ("home"). The changes in population mobility after social distancing differ across the 6 GMI categories, with retail, transit, and work categories being impacted the most overall. Because the GMI time series show strong weekly seasonality, we reduced the seasonality by smoothing over a 7-day window (see Materials and Methods), as seen in Figure S1.b. Figure S1 and Table S1, show that population mobility in all countries but Italy experienced a noticeable decrease (with GMI transit mobility decreasing to 85 and below) between March 12 and March 19, 2020, demonstrating that the implementation of social distancing measures occurred almost simultaneously across the 14 countries. A noticeable mobility reduction in Italy occurred about 2 weeks earlier, on February 24, 2020. In all countries but Italy, once GMI transit decreased below 85, it decreased continuously until it reached the minimum, and then stayed near the minimum for the remainder of our observation period on April 29. The lowest smoothed GMI transit mobility ranged from 13.9 in Spain to 56.2 in Sweden (Table S1, GMI Min column). In Italy, we can observe 2 distinct social distancing phases. Phase 1 social distancing measures resulted in GMI transit mobility between 70 and 85, which persisted for 14 days, from February 24 to March 8. After the stronger stay-at-home measures enforced by the government [27, 28], the smoothed GMI transit mobility decreased below 25 on March 15 and stayed below 25 (we refer to it as Phase 2 later in the paper).

Figure S2 shows COVID-19 death counts and death growth rates (in log-scale) for each country. As seen from the upper panels, the originally reported daily death counts are very noisy with outliers and weekly seasonality. The strongest weekly seasonality in the data is observed for Germany, Great Britain, Netherlands, and Sweden, which follow a pattern of evident dips in reported deaths during weekends, followed by subsequent jumps in reported deaths that compensate for the dips. In addition to seasonality, several countries such as France and United States have days with significant outliers. For example, during the period of April 3-6, 2020, the number of reported deaths in France is the following sequence: 471, 2,004, 1,053, and 518, indicating reporting inconsistencies. In addition, Sweden has several distinct underreporting periods, the largest surrounding the Easter holiday (April 12-14, 2020), that are not followed by compensatory upticks in counts. Because of the lack of consistent information about reporting issues in 14 countries, we developed a country-agnostic data preprocessing approach (see Materials and Methods). The preprocessed daily counts (Figure S2, middle panels) capture the overall trend and enable us to obtain a more reliable estimate of death growth rates (Figure S2, bottom panels). It is evident from Figure S2 and Table S1 that all 14 countries experienced a sizeable decrease in death growth rates during the observed period, from the highs in range from 0.144 (Turkey) to 0.259 (Italy) at the start of the outbreak to the lows in range from –0.043 (Spain) to 0.058 (Brazil) by second half of April. For our analysis, we excluded all days when the corrected plus smoothed death counts were below 10 because they cannot reliably estimate death growth rate, and the last 3 days due to the effects of smoothing.

**Regression Modeling of COVID-19 Growth Rate**

The objective was to use the collected data to fit a regression function that predicts the death growth rate from GMI mobility in all 14 countries. Using the preprocessed GMI and death count time series, we created a separate data set for each of the 14 individual countries. We created a different data set for each choice of day-delay $\Delta \in [1,30]$. Given $\Delta$, we formed data points $(m(t), \lambda_d(t + \Delta))$ for each country. Mobility was calculated as $m(t) = GMI(t)/100$, where $GMI(t)$ is one of the 6 GMI categories. For each data point, we also recorded the variance of $\lambda_d(t)$, needed to provide uncertainty weights for the regression, and daily change in population mobility, $|m(t + 1) - m(t)|$, needed to decide which data points to remove (see Materials and Methods).

In the first experiment, we assumed that the same delay $\Delta$ is appropriate for all countries. Thus, for each choice of delay $\Delta$ we joined data points from all 14 countries. Depending on $\Delta$, the total number of created data points ranged from 622 to 705. To fit a regression function $f$ that predicts $\lambda_d(t + \Delta)$ from $m(t)$, we had to decide what is the appropriate delay $\Delta$, which of the 6 GMI categories to use for population mobility $m(t)$, which order $k$ of polynomial is appropriate for the regression function $f$, and what threshold $\theta$ to use to remove data points with $|m(t + 1) - m(t)| > \theta$. Using a grid search, we determined that the best choices are $\Delta = 20$ days, $m(t) = GMI_{transit}(t)/100$, $k = 2$, and $\theta = 0.05$. The resulting regression model trained using Weighted Least Squares (WLS) is

$$\lambda_d(t + 20) = \alpha_0 + \alpha_2 m^2(t) = -0.042 + 0.214 \cdot m^2(t),$$

and its accuracy is $R^2 = 0.784$ (the adjusted coefficient of determination, which measures the fraction of the variance in $\lambda_d$ explained by $m$; $R^2 = 1$ is the theoretical maximum). We refer to this model as CovTM2 (Covid-19 growth rate model with Transit GMI with order-2 polynomial without the linear term).

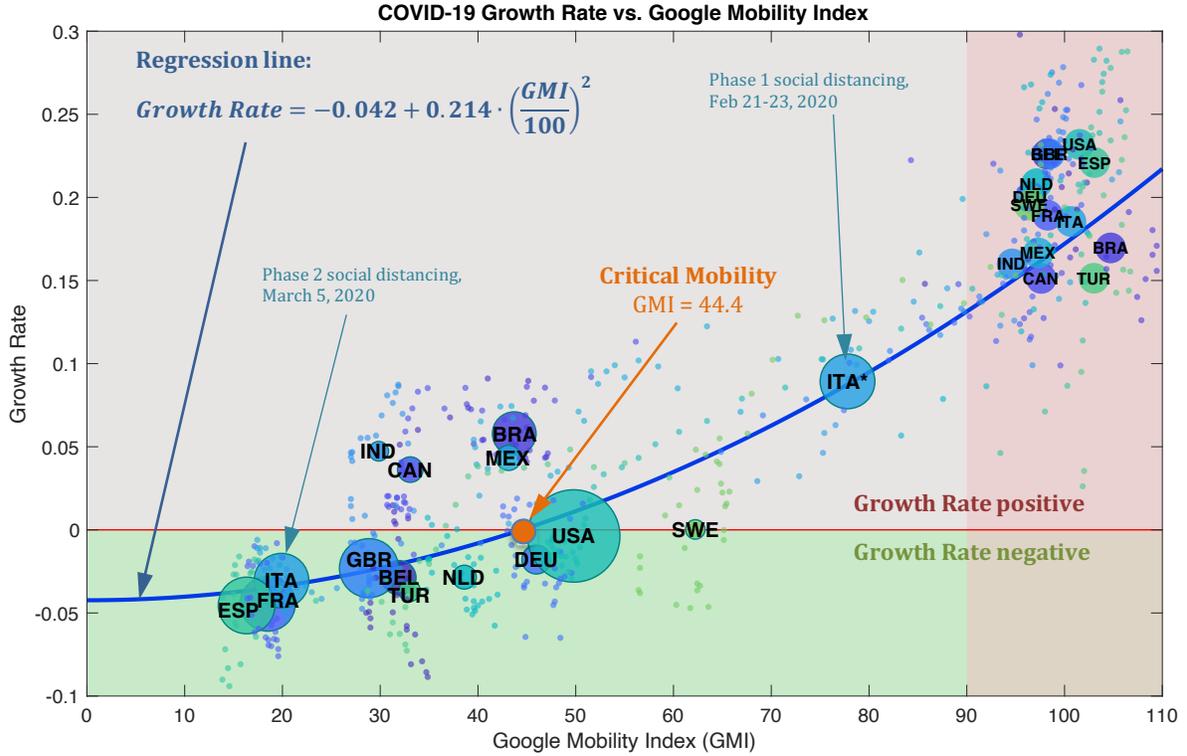

**Figure 2. CovTM2 Regression Model: Relationship between COVID-19 infection growth rate and population mobility in 14 countries.** The horizontal axis is the Google Mobility Index (GMI) for the transit category. A value of 100 corresponds to the baseline mobility in each country before social distancing. The vertical axis corresponds to the COVID-19 infection growth rate. Positive values represent a growing epidemic with an exponentially increasing number of cases (growth rate of 0.15 means 15% daily increase in number of cases), and negative values correspond to a declining epidemic. The blue curve is the regression function that implies a *quadratic relationship* (formula in the top left corner) between the GMI and the growth rate. The GMI value at the point where the regression function equals zero (blue line crosses the red line; GMI = 44.4%, 95% CI = [36.1,47.6]%) is the *critical mobility*, which is the estimated level of mobility necessary to stop the growth of the epidemic. Each small dot is a data point from one of the 14 countries from which CovTM2 is trained. All countries are represented by 2 circles (except for Italy, which is represented by 3 circles). The location of the first circle (on the far right) equals the average $GMI(t)$ and average COVID-19 death growth rate $\lambda_d(t+20)$ for days before social distancing, $(GMI(t) > 90$. The location of the second circle (on the left) equals the average $GMI(t)$ and average $\lambda_d(t+20)$ after social distancing took place ($GMI < \min(GMI) + 10$). The size of the circles on the left is proportional to the maximum number of daily deaths for a given country. For Italy, in addition to the circle on the right, there are two circles to the left: the one labeled ITA* corresponds to the first phase of social distancing measures announced from February 21-23, 2020; the one labeled ITA corresponds to the second phase of social distancing measures announced on March 5, 2020.

In Figure 2 we show the CovTM2 model with superimposed additional country-specific information to help interpret the results. CovTM2 accurately predicts $\lambda_d(t+20)$ from $m(t)$ for 9 of the 14 countries, namely Italy, France, Spain, UK, Belgium, Turkey, Netherlands, Germany and USA. Moreover, the results for Italy reveal that CovTM2 accurately predicts both phases of social distancing in Italy. Of the remaining 5 countries, Sweden's growth rate is lower than predicted from transit GMI, while Brazil, Mexico, India, and Canada have growth rates higher than predicted by the model. A visual inspection of Figure 2 and the high accuracy of $R^2 = 0.784$ indicate that GMI transit is a good proxy for measuring the effect of social distancing on the growth rate. It also shows that each country may have additional intricacies influencing the COVID-19 growth that are not fully captured by the CovTM2 mobility model.

From CovTM2, assuming that our data covers the initial stages of the epidemic ($S/N \approx 1$), we can estimate (see Materials and Methods): a) that the recovery rate in the SIR model is $\gamma = -\alpha_0 = 0.042$ 1/day (95% Confidence Interval (CI) [0.027, 0.054] 1/day), indicating that the expected time to recover from COVID-19 is $\gamma^{-1} = 23.8$ days; b) that the transmission rate at the baseline mobility level is $\beta_0 = \alpha_2 = 0.214$ 1/day (95% CI [0.186, 0.244] 1/day), indicating that the expected time needed for an infected person to infect another person is $\beta_0^{-1} = 4.67$ days, c) that the baseline reproduction number is $R_0 = \beta_0/\gamma = 5.06$ (95% CI [4.38, 7.19]), which is the expected number of people infected by a single infected person, and d) that the critical mobility equals $m_c = \sqrt{-\alpha_0/\alpha_2} = 0.444$ (95% CI [0.361, 0.476]). The confidence intervals are bootstrap estimate (see Materials and Methods) based on 1,000 replicates (see Figure S3). The optimal delay $\Delta = 20$ days is relatively close to the recovery time $\gamma^{-1} = 23.8$ days, even though the regression model did not impose it. All those estimates are comparable to what is currently empirically known about the transmission dynamics and progression of the COVID-19 disease [29-32]. These results indicate that population mobility needs to be reduced to less than half of its baseline value to be able to start reducing the growth of COVID-19 epidemic.

To demonstrate that the choice of parameters for CovTM2 ($\Delta$, GMI category, polynomial order, and $\theta$) is stable we show the sensitivity analysis results. Figure S4 shows the $R^2$ accuracy when the delay $\Delta$ ranges from 1 to 30 days for each of 6 different categories of GMI, while keeping $k = 2$ and $\theta = 0.05$. The accuracy for all 6 categories changes smoothly over the range of delays. The work category of GMI achieves a comparable accuracy to CovTM2 of $R^2 = 0.773$ for $\Delta = 18$ days, while the accuracy of other categories is substantially lower, with the park category being the least predictive (peaking at $R^2 = 0.326$ when $\Delta = 25$ days). Figure S5 shows changes in accuracy as a function of the $\theta$ threshold for data removal. It could be observed that $\theta$ has a minor influence on the regression model. This is a strong indicator that the rate of change in population mobility does not have a pronounced negative effect on modeling, although our analysis (see Materials and Methods) implied it could be an important factor. We decided to use a removal threshold $\theta = 0.05$ for CovTM2 because it results in a slightly higher accuracy than other choices. Table S2 shows the fitted regression models for 5 different polynomials and their accuracies when other parameters are the same as CovTM2. Figure S6 depicts the 5 learned regression models on a scatterplot. As can be seen, the first order polynomial is inferior, while 2nd and 3rd order polynomials have the same accuracy. The 2nd order polynomial has a negative linear term and is therefore not an increasing function of $m(t)$ in the whole range from 0 to 1. Thus, we also fitted the second order polynomial without the linear term, which is the best-fit quadratic polynomial that is constrained to be an increasing function of $m(t)$ in the whole range. This constrained second-order polynomial is used in CovTM2.

In the second set of experiments, we explored a country-specific relationship between mobility and growth rate. In order to reduce the degrees of freedom in polynomial fitting, we set the constraint that each country-specific order-2 regression function has an identical intercept $\alpha_0$, by assuming that the recovery rate $\gamma = -\alpha_0$ is a general property of the epidemic and should not be country-dependent. We also noticed that the average post-social distancing growth rate in France and Spain is slightly lower than the intercept of CovTM2. Thus, we set $\alpha_0$ for all country-specific models to $-0.05$, slightly lower than that of CovTM2 (but, still within its 95% CI). In addition, we also allowed for the delay $\Delta$ to be country-dependent, as long as it remained in the range from 15 to 25 days, to remain close to the overall optimum $\Delta = 20$ days. We selected the $\Delta$ for each country that results in the highest predictive accuracy.

We show the results for Italy in Figure 3 and for all 14 countries in Figure S7 and Table S3. As expected, for the 9 countries that were already a good fit for the CovTM2 model, their country-specific model is very similar to CovTM2. For Italy, in particular, we can observe that after adjusting the delay to $\Delta = 15$ days the fitting becomes highly accurate ($R^2=0.984$) and closely follows both phases of population mobility reduction. For Brazil, Canada, and Mexico, their models are almost linear, while for India the resulting model has a concave shape. The reported country-specific $R^2$ accuracies are very high for all 14 countries, ranging from 0.755 to 0.984 with the median of 0.903.

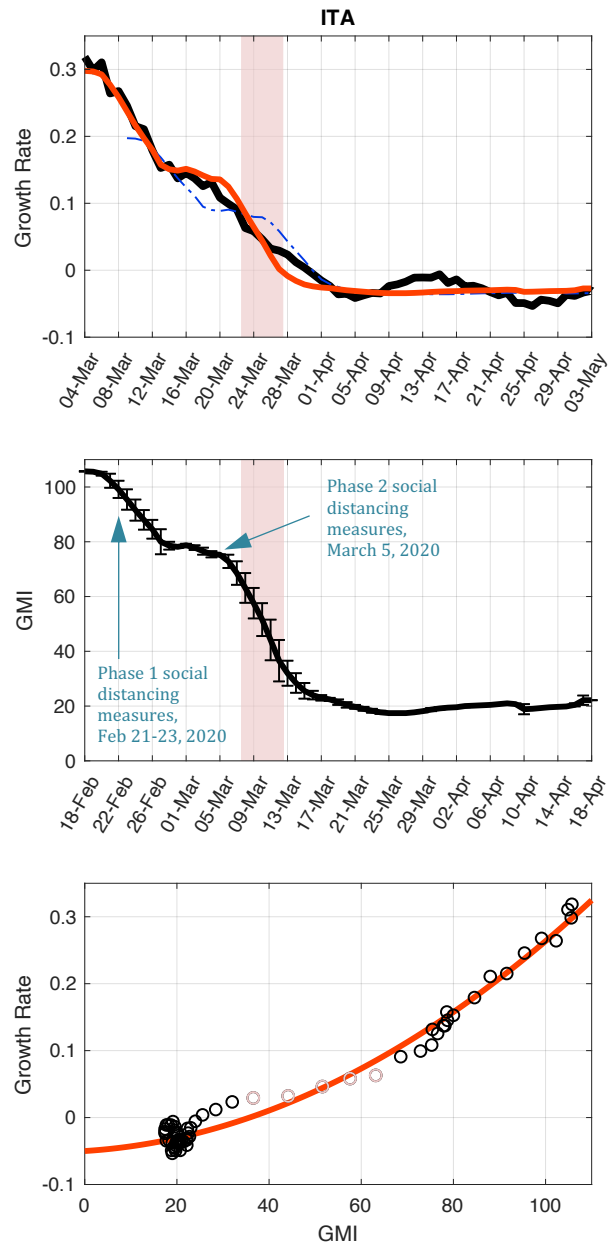

**Figure 3. Country-specific regression model for Italy**. The top panel shows the observed growth rate (black curve), the country-specific regression model (orange curve), and the CovTM2 model (dashed blue line). The part shaded in light red shows where data points were not used in the fitting because the mobility change is above the threshold. The middle panel shows the GMI transit time series. Note that it is shifted back in time relatively to the growth rate (top panel) by a country-specific delay Δ= 15 days. The bottom panel shows the learned relationship between GMI transit and growth rate (orange curve) that corresponds to the orange curve in the top panel, and the scatter plot of data points used for model fitting (black dots) as well as the removed points (light orange dots).

In addition to the GMI transit category, we also trained country-specific models with other GMI categories. In Table S3 we list the highest obtained $R^2$ with the corresponding GMI category and delay. Interestingly, for 11 out of 14 countries, GMI transit is the best or very close to the best category. The only exceptions are Belgium (GMI retail increases $R^2$ by 0.064), Mexico (GMI work increases $R^2$ by 0.074) and Turkey (GMI parks increases $R^2$ by 0.049).

By using the country-specific delays from Table S3, we created a data set for each country using its best delay, and then joined those data sets to create a delay-optimized data set with 642 data points, using GMI transit mobility as $m(t)$. The resulting regression model trained on this data set was

$$\lambda_d(t + \Delta) = \alpha_0 + \alpha_2 m^2(t) = -0.044 + 0.242 \cdot m^2(t),$$

where 95% CI for $-\alpha_0$ is [0.029,0.057] and for $\alpha_2$ [0.217,0.268]. The accuracy of the model is $R^2 = 0.801$, higher than that of CovTM2 model. Its critical mobility is 0.425 (with 95% CI [0.349,0.466]). We refer to this model as CovTM2d (with country-specific delays) and this model is illustrated in Figure S8.

## Discussion

Despite the widespread implementation of social distancing measures across the globe to contain the COVID-19 pandemic, its quantitative relationship with the epidemic growth rate has remained an open question. The results of this study indicate that there is a quantitative relationship between the population mobility, measured by the Google Mobility Index (GMI), and the growth rate of COVID-19 infections, namely: the growth rate is a quadratic function of the mobility index. This function was obtained via a regression model trained on publicly available data from 14 countries, and it provides a direct estimate of the important parameters of the epidemic such as the mobility-dependent transmission and mobility-independent recovery rates, as well as the critical mobility. An important finding of this study is that a single function that relates mobility and infection growth fitted on 14 countries achieves high accuracy. Given the significant differences in combination of social distancing measures introduced across these countries, as well as underlying social, demographic and cultural differences, the quality of model fit is very remarkable.

The regression model was derived from first principles encoded in an SIR compartmental epidemiological model, which is one of the simplest models that can still generate a dynamic that qualitatively resembles an epidemic such as COVID-19. In fact, thanks to the SIR simplicity and augmenting it by the mild assumptions needed to model the death growth rate, the resulting regression model is easy to interpret, easy to fit, and accurate, thus hitting a sweet spot between modeling simplicity and accuracy. There is a range of extensions of the basic SIR model that can generate more realistic behaviors, such introducing more compartments to model the effects of incubation or stages of an infection, or using heterogeneously mixing populations to account for age-dependent and spatial effects. However, more complex epidemiological models have a larger number of parameters that would necessitate more complex data and result in a less interpretable statistical model. Our study relied exclusively on population mobility as a proxy for changed behavior due to social distancing. Using additional population behavior (such as data about mask usage) and environmental (such as weather) variables would increase the expressive power of our statistical model. However, we are currently not aware of other high quality behavior variables with a global coverage that could have been used in our study. Due to the small size of our data, we reason that environmental time series (although they are publicly available and of good quality) could lead to overfitting of our model.

Our results imply that the COVID-19 transmission rate is a quadratic function of the GMI transit mobility. While understanding why is a topic for future research, a quadratic function would arise if we assumed that the population is homogeneous (as is the case in the SIR model) and that the sole impact of social distancing is in reduction of number of contacts between infected and susceptible people (infection probability per each contact does not change). In this case, the transmission rate would scale with the number of contacts, which would scale as the square of population mobility if the fraction of infected people is small.

The quadratic relationship implies that while an initial moderate mobility reduction would lead to a sizeable reduction in COVID-19 infection growth rate, any additional mobility reduction would have diminishing returns. Another undesirable implication of the quadratic relationship between population mobility and infection growth rate is that, starting from a given level of reduced population mobility, any increase in mobility will have a much larger negative effect (rapid increase in growth rate) than any decrease in mobility (mild reduction in growth rate). Reversing the growth of an epidemic requires a significant reduction in mobility (critical mobility below 50% of the baseline pre-COVID-19 mobility). To achieve a robust decline in infections and deaths, population mobility would need to go much lower and become comparable to the reductions achieved through strong social distancing measures in Italy, Spain, and France.

Overall, our results imply that any attempt to eradicate the epidemic before much of the population gets infected and in the absence of pharmaceutical interventions, would require disciplined, coordinated, and prolonged social distancing. Another option would be to allow a sizeable fraction of the population to get infected, in what case the susceptible ratio $\mu = S/N$ would be significantly below 1. Because in the SIR model the transmission rate is multiplied by the susceptible ratio when calculating the infection growth rate, this would lead to a reduction in the infection growth rate even if the population mobility is kept constant.

It is becoming evident that population mobility reduction due to social distancing is an overwhelmingly costly intervention. Thus, it is of utmost importance to implement less crude NPIs or to combine social distancing with other forms of NPIs. Our results on individual countries indicate that population mobility reduction is not the only effective control mechanism. First, it is evident that baseline COVID-19 growth rate prior to social distancing differed across countries (before social distancing the daily growth rate differed from 14.4% to 25.9%). Moreover, growth rates observed after introduction of social distancing show that mobility reduction did not have the same effect in all countries. Five countries, in particular, were outliers with respect to the learned regression function (Sweden, Canada, Brazil, Mexico, and India) and they might offer useful insights about COVID-19 control. Probably the most important research objective outside of finding a cure or vaccine will be related to *precision social distancing*: to understand how governments and public health authorities could control or modify select aspects of society and how individual people should fine-tune their behavior in a way that minimally impacts the economy and lifestyles while at the same time curbing or reversing the growth of the epidemic.

## Materials and Methods

For a time-dependent function $x(t)$, its *rate of change* is defined as time derivative $x'(t) = dx(t)/dt$ and its (relative) *growth rate* is defined as $\lambda_x(t) = x'(t)/x(t)$. In this study we assume that the unit of time is one day. If the function changes slowly within a unit of time (a day), as is the case for the functions relevant for the COVID-19 epidemic, we can approximate the rate of change via finite differences of daily changes as $x(t+1) - x(t)$, and the growth rate via $(x(t+1) - x(t))/x(t)$ or via $\log(x(t+1)/x(t))$.

Our framework for modeling the impact of social distancing on the COVID-19 epidemic growth, outlined in Figure 1, depends on three models described below (epidemiological model, social distancing model, and statistical model).

### Epidemiological Model of COVID-19 Infection Growth

An established way to capture the basic dynamics of infectious diseases such as COVID-19 is via compartmental epidemiological models. We start from the basic Susceptible-Infectious-Recovered (SIR) model,

$$\frac{dS}{dt} = -\frac{\beta IS}{N}, \qquad \frac{dI}{dt} = \frac{\beta S}{N}I - \gamma I, \qquad \frac{dR}{dt} = \gamma I, \qquad (1)$$

which describes the temporal evolution of the numbers of susceptible, *S(t)*, infected, *I(t)*, and recovered, *R(t)*, individuals at time *t* from a population of size *N*, such that *S(t) + I(t) + R(t) = N*. The parameter $\beta$ is the transmission rate and $\gamma$ is the recovery rate. We assume that $\beta = \beta(t)$ is time-dependent due to its dependence on social distancing (see below), while γ remains constant (in the absence of a cure).

In this work, we are studying COVID-19 during the initial stages of epidemic, when a large majority of people are still susceptible. Thus, we make a simplifying assumption that $S(t)/N$ is sufficiently close to 1 and that $\lambda_S(t) = S'(t)/S(t)$ is sufficiently close to 0 (because $I(t)/N$ is close to 0) during the observed period. With that, the SIR model (1) reduces to the need to track *I(t)*, via the equation

$$I'(t) = \beta(t)I(t) - \gamma I(t) = \lambda_I(t)I(t), \qquad (2)$$

where the infection growth rate equals $\lambda_I(t) = I'(t)/I(t) = \beta(t) - \gamma$. From this equation, we obtain the basic reproduction number $R_0 = \beta/\gamma$, which varies in time as $\beta$ changes. From (2) we also obtain the number of new infections per day happening at time $t$, as $j(t) = \beta(t)I(t)$.

A key shortcoming of ordinary differential equation models like (1) is that they do not explicitly capture delay and temporal non-locality effects. While this can be justified for the infection dynamics, for COVID-19 the time between infection and death is usually weeks, and thus the simple $R(t)$ dynamics in the SIR model do not suffice. We therefore retain SIR as a simplified model of COVID-19 infections, but use a more realistic model for the dynamics of COVID-19 deaths. Let $d(t)$ denote the number of new deaths per day happening at time $t$. We assume that the probability of death for a newly infected individual is $p_d$ and that the time of death is distributed over some future time, which gives rise to the following law

$$d(t + \Delta) = p_d \int G(s)j(t + s)ds = p_d \int G(s)\beta(t + s)I(t + s)ds. \qquad (3)$$

Here the shift parameter $\Delta$ is the average number of days between infection and death (if death occurs) and $G$ is a kernel (normalized, $\int G(s)ds = 1$, and centered, $\int sG(s)ds = 0$) describing the distribution of times of deaths around the delay $\Delta$.

Based on the infection law (2) and death law (3), we obtain for the death growth rate

$$\begin{aligned}\lambda_d(t + \Delta) &= \frac{d'}{d}(t + \Delta) = \frac{\int G(s)\beta(t + s)I'(t + s)ds + \int G(s)I(t + s)\beta'(t + s)ds}{\int G(s)\beta(t + s)I(t + s)ds} \\ &= \frac{\int G(s)\beta(t + s)(\lambda_I(t) + \lambda_I'(t)s + \cdots)I(t + s)ds}{\int G(s)\beta(t + s)I(t + s)ds} + \frac{\int G(s)I(t + s)\beta'(t + s)ds}{\int G(s)I(t + s)\beta(t + s)ds} \\ &= \lambda_I(t) + \beta'(t)\frac{\int sG(s)\beta(t + s)I(t + s)ds}{\int G(s)\beta(t + s)I(t + s)ds} + \cdots + \frac{\int G(s)I(t + s)\beta'(t + s)ds}{\int G(s)I(t + s)\beta(t + s)ds}.\end{aligned} \qquad (4)$$

In the special case when $\beta(t)$ is constant in time around time $t$, i.e., $\beta'(t) = 0$, this relationship substantially simplifies to $\lambda_d(t + \Delta) = \lambda_I(t)$, which provides a direct expression for death growth rate as a time-delayed function of infection growth rate $\lambda_I(t)$. In addition, under the same assumption that $\beta'(t) = 0$, it also follows that $\lambda_j(t) = \lambda_I(t)$, such that we can write

$$\lambda_I(t) = \lambda_j(t) = \lambda_d(t + \Delta). \qquad (5)$$

It must be stressed that this relationship (5) does not require any knowledge of the virus's lethality $p_d$.

In turn, when $\beta$ and $\lambda$ are varying in time (i.e., immediately after the implementation of social distancing measures), there is no simple relationship between death growth rate and infection growth rate (the additional integral terms in (4) do not vanish). Thus, the data from periods during which $\beta$ experiences rapid change will be ignored in our statistical model, as will be described later in this section.

**Social Distancing Model: Infection Growth Rate as a Function of Population Mobility**

The transmission rate $\beta(t)$ in the reduced SIR model (2) depends on the population behavior $b(t)$ at time $t$. However, it is not clear at the moment what population behavior variables would be sufficient to precisely predict $\beta(t)$ and how those variables could be reliably measured over many countries. Instead of using population behavior, in this study we rely on the Google Mobility Index (GMI) [9], which is a publicly available data set about population mobility in a large number of countries calculated using a standardized methodology. We hypothesize that GMI can be used as a proxy for $b(t)$, because population mobility reflects some aspects of population behavior change in response to social distancing. Specifically, to obtain $m(t)$ we use the publicly available Google Mobility Index (GMI) [9]. The $GMI(t)$ variable is created by adding 100 to the original GMI values and can be interpreted as the percent of the baseline mobility. Thus, $GMI(t) = 100$ represents the baseline mobility and $GMI(t) = 0$ refers to the hypothetical case when the population stops moving completely. We define the mobility variable as $m(t) = GMI(t)/100$, such that it measures the fraction of the baseline mobility.

Our main hypothesis is that the transmission rate $\beta$ can be statistically described as a function of population mobility, i.e. $\beta(t) = g(m(t))$. We denote $\beta_0 = g(1)$ as the baseline transmission rate. We constrain $g$ to be a non-decreasing function of $m$ for $0 < m < 1$. Moreover, we require that $g(0) = 0$, reflecting the assumption that the infection transmission halts in the theoretical limit when the population stops moving. Thus, we model $g(m)$ as a polynomial with zero intercept. Hence, we obtain the infection growth rate $\lambda_I(t) = f(m(t))$, where $f(m) = g(m) - \gamma$.

For the data fitting, we choose polynomials of order $k$, i.e., $\lambda_I(t) = f_k(m(t))$, where $k$ is a hyperparameter that is determined experimentally. Moreover, we constrain the function $f_k(m)$ to be non-decreasing on the interval $0 < m < 1$, as there is no principled reason why an increase in mobility should ever result in a decrease in transmission rate.

**Statistical Model: Predicting COVID-19 Growth Rate from Population Mobility**

By combining the social distancing model and the SIR epidemiological model, the resulting relationship between population mobility and death growth rate is

$$\lambda_d(t + \Delta) = f_k(m(t)). \tag{6}$$

It states that death growth rate at time $t + \Delta$ is a function of mobility at time $t$. We note that both $m(t)$ and $\lambda_d(t)$ are measurable and publicly available, unlike other quantities such as $b(t), \beta(t), \gamma$, that are not easy to measure, or $I(t), S(t)$, and $R(t)$, that are unreliable. Thus, it will be possible to fit the function $f_k$ using observed data about population mobility and death growth rate.

Once a functional relationship between death growth rate and mobility is learned from data, from (5) we also know the infection growth rate as $\lambda_I(t) = f_k(m(t))$. Moreover, the recovery rate $\gamma$ and the transmission rate $\beta(t)$ can also be inferred from $f_k$. The root of the function $f_k(m)$ determines the *critical mobility* $m_c$ (where $f_k(m_c) = 0$) below which the infection decays.

**Data Preprocessing**

The Google Mobility Index (GMI) exhibits a strong weekly seasonality, where mobility during weekdays is noticeably different from weekends. Because the kernel $G$ spreads deaths around $\Delta$ days after infection, weekly seasonality in mobility becomes a source of covariate noise in our statistical model. To reduce this noise we remove seasonality by smoothing GMI with moving averaging over 7 days, $GMI_{smooth}(t) = \sum_{j=-3}^{3} GMI(t + j)$. In our experiments, we use $m(t) = GMI_{smooth}(t)/100$. We note that it has been estimated that COVID-19 deaths occur in a range from 2 to 8 weeks after infection, but the actual time distribution is not known. Thus, the 7-day smoothing of mobility should not have an adverse effect on the growth rate prediction.

Daily death counts $d(t)$ are very noisy (see Figure S2, top subplot) and require data cleaning and processing. Our approach has three steps: (Step 1) outlier removal, (Step 2) undercount/overcount correction, (Step 3) smoothing. For outlier correction (Step 1), we manually corrected only 10 entries. We replaced a single zero entry occurring in Germany, Spain, Sweden, and U.S. by an entry from the previous day. We clipped two U.S. entries over 3,000 to 3,000 and four France entries over 1,000 to 1,000.

For undercount/overcount correction (Step 2), we allowed transferring some overcount deaths to previous days if they had an undercount. Because deaths are sampled with probability $p$ from the newly infected, we treat $d(t)$ as the binomial random variable. When $d(t)$ is sufficiently large, it could be approximated as a normal random variable with mean $d(t)$ and variance $d(t)$. By recognizing that the growth in deaths is exponential, we first calculate the local geometric mean for deaths, $\tilde{d}(t)$, as $\log\left(\tilde{d}(t)\right) = \frac{1}{7}\sum_{j=-3}^{3}\log\left(d(t+j)\right)$. If at any day $t$, $d(t)$ is sufficiently larger than $\tilde{d}(t)$, $d(t) - \tilde{d}(t) > w\sqrt{d(t)}$, where $w$ is a constant ($w = 0.2$), we consider the excess deaths, $e(t) = d(t) - \tilde{d}(t) - w\sqrt{\tilde{d}(t)}$, as the overcount. Then, we look at the previous day, and declare undercount if $\tilde{d}(t-1) - d(t-1) > w\sqrt{\tilde{d}(t-1)}$. In this case, we transfer up to $e(t)$ deaths to $d(t-1)$, as long as $d(t-1)$ remains below $\tilde{d}(t-1) - w\sqrt{\tilde{d}(t-1)}$. If there is any budget left in $e(t)$, we consider transfer of deaths to previous days, $d(t-2)$, $d(t-3)$, and $d(t-4)$. The same correction procedure is repeated for every overcount day. Then, the entire process is repeated a several times until convergence (no more reassignments of deaths). After the convergence, we compare the difference $d(t) - \tilde{d}(t)$ for each day, and, if it is beyond one standard deviation, $\sqrt{\tilde{d}(t)}$, we set $d(t)$ at one standard deviation from $\tilde{d}(t)$.

After the correction (Step 2), the time series $d(t)$ will still have seasonality effects. Thus, in Step 3, we calculate its local geometric mean $\tilde{d}(t)$ over a moving window of size 7 (see Figure S2, middle subplot) and use it to calculate the death growth rate as $\lambda_d(t) = \log\frac{\tilde{d}(t+1)}{\tilde{d}(t)} = \frac{1}{7}\log\frac{d(t+4)}{d(t-3)}$. Thus, $\lambda_d(t)$ is a log-ratio of two binomial variables, which can be approximated [33] as a normal random variable with mean $\lambda_d(t)$ and variance $\frac{1}{7^2}\left(\frac{1-p}{d(t+4)} + \frac{1-p}{d(t-3)}\right)$, where $p$ can be neglected because it is small. Eventually, we removed $\lambda_d(t)$ on days when $\tilde{d}(t) < 10$ and the last 3 days (because of the smoothing).

**Regression Model for COVID-19 Growth Rate**

Given a time series of preprocessed daily mobility indices, $\{m(t)\}$, death growth rates, $\{\lambda_d(t)\}$, and assuming a second-order polynomial, we can fit the following regression function,

$$\lambda_d(t+\Delta) = f_k(m(t)) = \alpha_0 + \alpha_1 m(t) + \alpha_2 m(t)^2. \tag{7}$$

The parameters $\alpha_0$, $\alpha_1$, and $\alpha_2$ can be fitted using least squares and $\Delta$ is a hyperparameter that can be obtained by line search. We impose a constraint that $f_k$ is an increasing function of $m$ in range $0 < m < 1$. Assuming $\Delta$ is known, the data set for fitting consists of matched pairs $(m(t), \lambda_d(t+\Delta))$. Because equation (5) is correct only when $\beta(t)$ does not vary too rapidly, we remove data points when $|m(t+1) - m(t)| > \theta$, where $\theta$ is the removal hyperparameter. Because $\lambda_d(t)$ is treated as a random variable with variance that depends on $d(t)$, to fit (7) we use Weighted Least Squares (WLS), where the weights in WLS are the reciprocals of the variance. To reduce overfitting we use feature selection to select the most appropriate GMI category as $m(t)$ and discard the others.

**Confidence Intervals by Bootstrapping**

Because our data set consists of time series from 14 countries, we use bootstrapping with two levels of sampling to provide confidence intervals for our regression model. For each bootstrap sample, we first select 14 countries randomly with resampling from the 14 countries. Then, for each sampled country we select data points randomly with resampling. We fit the regression models on 1,000 bootstrap samples and report 95% confidence intervals for regression parameters and critical mobility.

**Acknowledgements**
We thank Google for providing its population mobility data to public. We thank Saman Enayati, Ziyu Yang, and Tamara Katic for their help with data collection and processing.

# Supplementary Tables and Figures

**Table S1**. COVID-19 death and mobility statistics for 14 selected countries: Countries and their abbreviations, maximum recorded number of daily deaths (Count D Max), date of the maximum recorded number of deaths (Date D Max), date when number of daily deaths reached 10 (Date D ≥ 10), date when population mobility measured by GMI transit value decreased below 85 (Date GMI < 85%), minimum smoothed GMI transit mobility (GMI Min), average estimated death growth rate before social distancing corresponding to the days when GMI transit was over 90 ($\lambda_d(t)$ Before), and average death growth rate after social distancing corresponding to the days when GMI transit was within 10 of the GMI Min ($\lambda_d(t)$ After).

| Country | | Count D Max | Date D Max | Date D ≥ 10 | Date GMI < 85 | GMI Min | $\lambda_d(t)$ Before | $\lambda_d(t)$ After |
|---|---|---|---|---|---|---|---|---|
| Belgium | BEL | 496 | 4/11 | 3/21 | 3/13 | 28.5 | 0.214 | −0.036 |
| Brazil | BRA | 751 | 5/09 | 3/25 | 3/17 | 38.9 | 0.164 | 0.058 |
| Canada | CAN | 207 | 5/02 | 3/28 | 3/13 | 30.7 | 0.164 | 0.028 |
| France | FRA | 2,004 | 4/04 | 3/10 | 3/14 | 16.4 | 0.192 | −0.043 |
| Germany | DEU | 315 | 4/16 | 3/20 | 3/14 | 42.6 | 0.200 | −0.018 |
| Great Britain | GBR | 1,172 | 4/22 | 3/15 | 3/15 | 26.9 | 0.238 | −0.021 |
| India | IND | 195 | 5/05 | 4/02 | 3/19 | 27.0 | 0.158 | 0.048 |
| Italy | ITA | 971 | 3/28 | 3/03 | 2/24 | 17.4 | 0.259 | −0.029 |
| Mexico | MEX | 257 | 5/08 | 4/03 | 3/18 | 39.6 | 0.170 | 0.045 |
| Netherlands | NLD | 234 | 4/08 | 3/18 | 3/12 | 34.7 | 0.237 | −0.024 |
| Spain | ESP | 950 | 4/03 | 3/10 | 3/12 | 13.9 | 0.252 | −0.043 |
| Turkey | TUR | 127 | 4/20 | 3/22 | 3/16 | 26.9 | 0.144 | −0.042 |
| Sweden | SWE | 185 | 4/22 | 3/25 | 3/13 | 56.2 | 0.182 | −0.001 |
| United States | USA | 4,928 | 4/16 | 3/13 | 3/15 | 47.1 | 0.236 | −0.002 |

**Table S2.** Regression model accuracies for different orders of polynomial when GMI transit category is used for mobility $m(t)$. The last row corresponds to the second order polynomial with zero linear term, which guarantees that the growth rate is an increasing function of $m$. All numbers in the Model column have the unit 1/day

| Order | Model | $R^2$ |
|---|---|---|
| 1 | $-0.092 + 0.234 \cdot m$ | 0.732 |
| 2 | $-0.038 - 0.002 \cdot m + 0.231 \cdot m^2$ | 0.784 |
| 3 | $-0.049 + 0.069 \cdot m - 0.046 \cdot m^2 + 0.110 \cdot m^3$ | 0.784 |
| 2 (CovTM2) | $-0.042 + 0.214 \cdot m^2$ | 0.784 |
| 2 (CovTM2d) | $-0.044 + 0.242 \cdot m^2$ | 0.802 |

**Table S3.** Country-specific regression models. Columns 2-4 correspond to the best model for GMI transit category. Columns 5-8 (GMI Best) correspond to the best overall GMI category for each country. The numbers in the Model column have the unit 1/day, and the numbers in the Δ columns have the unit day.

| Country | GMI Transit | | | GMI Best | | |
|---|---|---|---|---|---|---|
| | Model | $R^2$ | Δ | Category | $R^2$ | Δ |
| BEL | $-0.05 + \phantom{0.000}0 \cdot m + 0.027 \cdot m^2$ | 0.829 | 22 | Retail | 0.893 | 22 |
| BRA | $-0.05 + 0.283 \cdot m - 0.008 \cdot m^2$ | 0.838 | 22 | Transit | 0.838 | 22 |
| CAN | $-0.05 + 0.265 \cdot m - 0.005 \cdot m^2$ | 0.832 | 25 | Work | 0.856 | 24 |
| FRA | $-0.05 + \phantom{0.000}0 \cdot m + 0.025 \cdot m^2$ | 0.941 | 19 | Transit | 0.941 | 19 |
| DEU | $-0.05 + \phantom{0.000}0 \cdot m + 0.025 \cdot m^2$ | 0.943 | 20 | Transit | 0.943 | 20 |
| GBR | $-0.05 + 0.030 \cdot m + 0.027 \cdot m^2$ | 0.984 | 17 | Transit | 0.984 | 17 |
| IND | $-0.05 + 0.375 \cdot m - 0.016 \cdot m^2$ | 0.755 | 18 | Grocery | 0.776 | 18 |
| ITA | $-0.05 + 0.036 \cdot m + 0.026 \cdot m^2$ | 0.977 | 15 | Transit | 0.977 | 15 |
| MEX | $-0.05 + 0.245 \cdot m - 0.001 \cdot m^2$ | 0.780 | 19 | Work | 0.854 | 18 |
| NLD | $-0.05 + \phantom{0.000}0 \cdot m + 0.029 \cdot m^2$ | 0.947 | 15 | Transit | 0.947 | 15 |
| USA | $-0.05 + \phantom{0.000}0 \cdot m + 0.027 \cdot m^2$ | 0.930 | 20 | Transit | 0.930 | 20 |
| ESP | $-0.05 + \phantom{0.000}0 \cdot m + 0.029 \cdot m^2$ | 0.953 | 15 | Parks | 0.959 | 16 |
| TUR | $-0.05 + 0.027 \cdot m + 0.016 \cdot m^2$ | 0.806 | 22 | Parks | 0.858 | 23 |
| SWE | $-0.05 + \phantom{0.000}0 \cdot m + 0.022 \cdot m^2$ | 0.876 | 22 | Work | 0.893 | 18 |

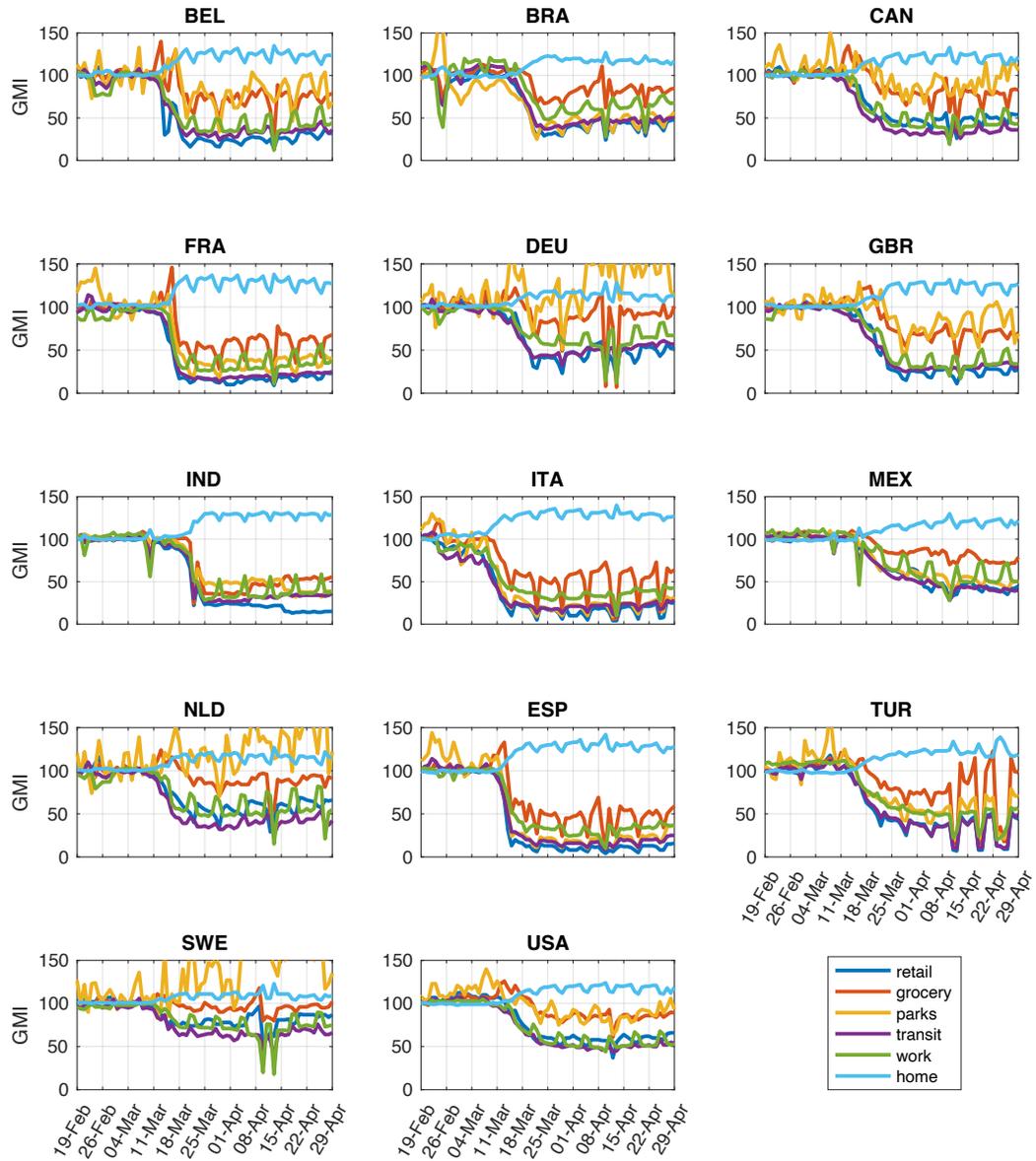

**Figure S1.a.** Google Mobility Index (GMI) time series for 6 mobility categories for 14 countries.

**Figure S1.b**. Smoothed Google Mobility Indices (GMI) time series for 6 mobility categories for 14 countries.

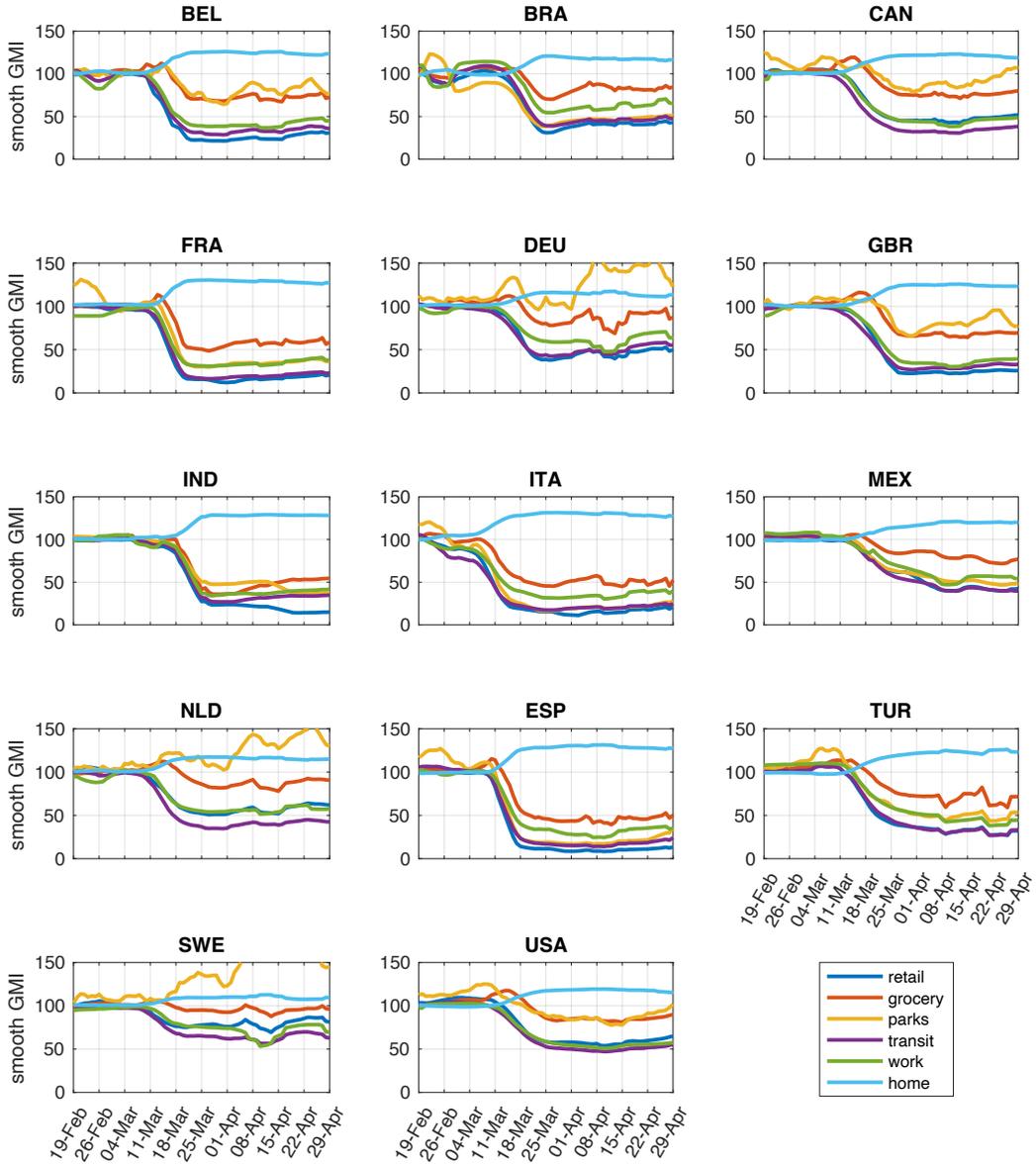

**Figure S2**. Daily deaths (top panels), preprocessed (corrected and smoothed) daily deaths (middle panels), and death growth rate $\lambda_d(t)$ derived from the preprocessed deaths (bottom panels) for 14 countries. The bottom panel also shows the uncertainty region of one standard deviation (shaded blue) for the growth rate.

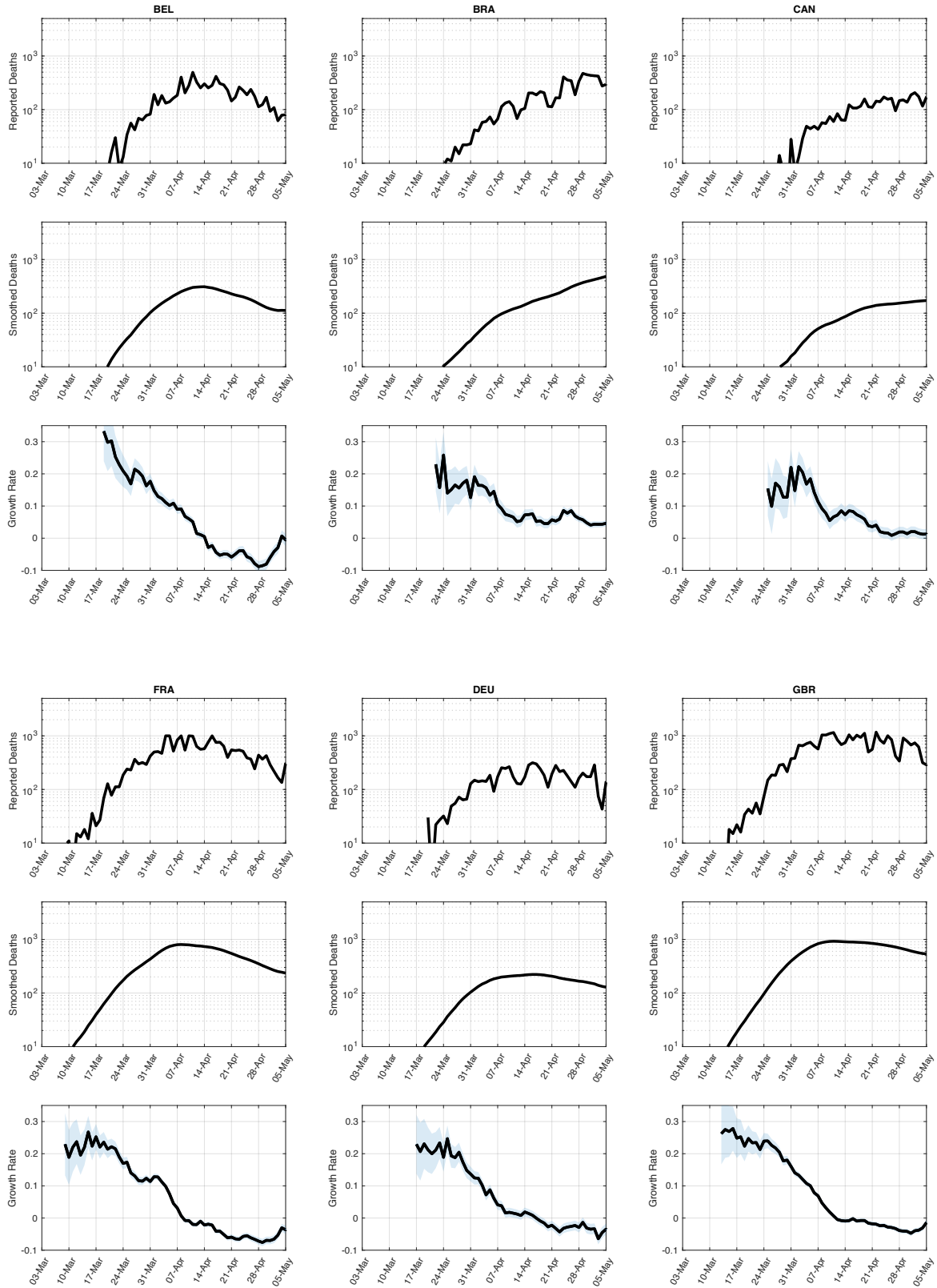

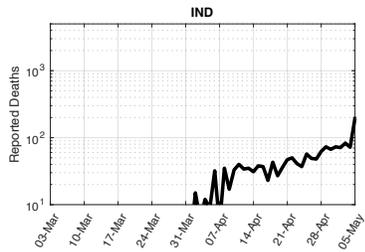
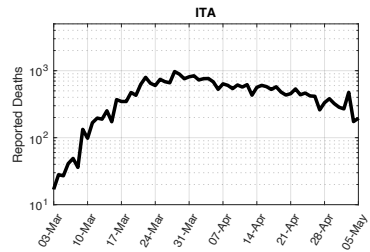
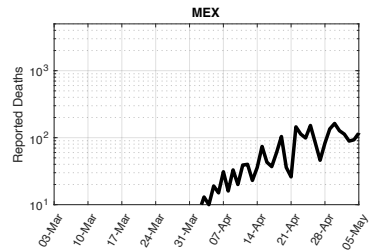
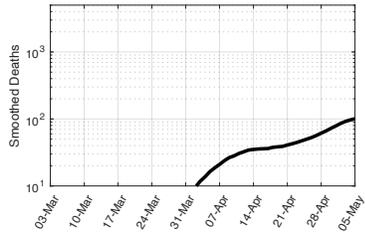
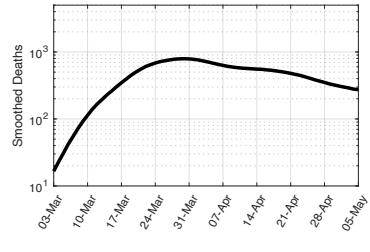
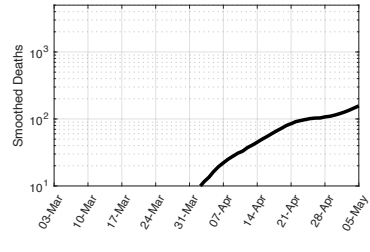
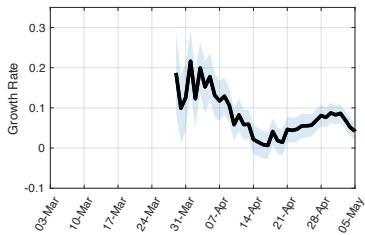
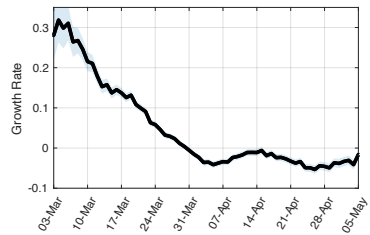
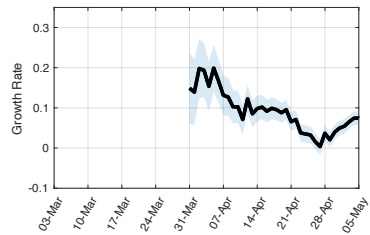
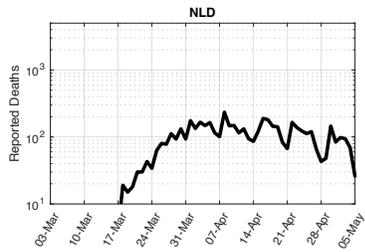
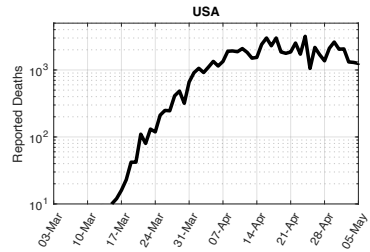
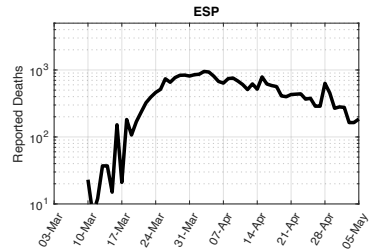
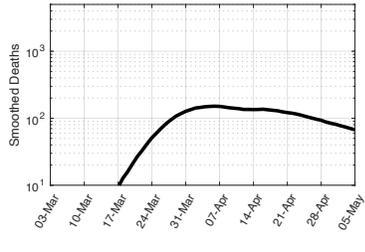
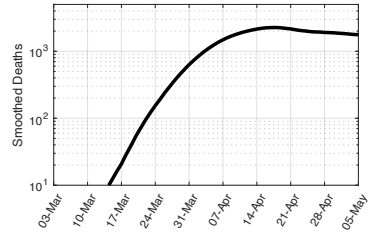
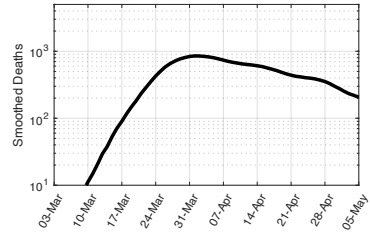
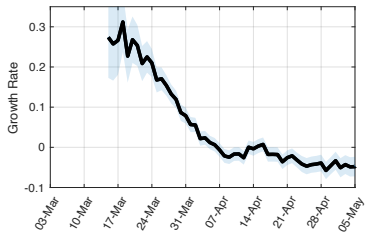
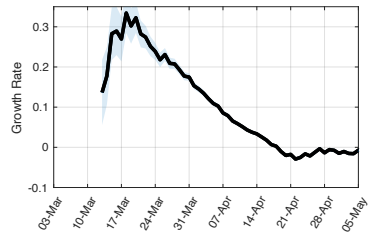
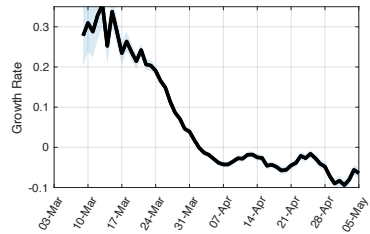

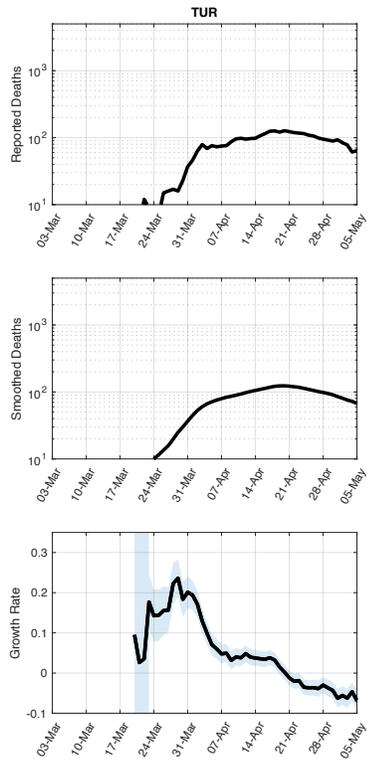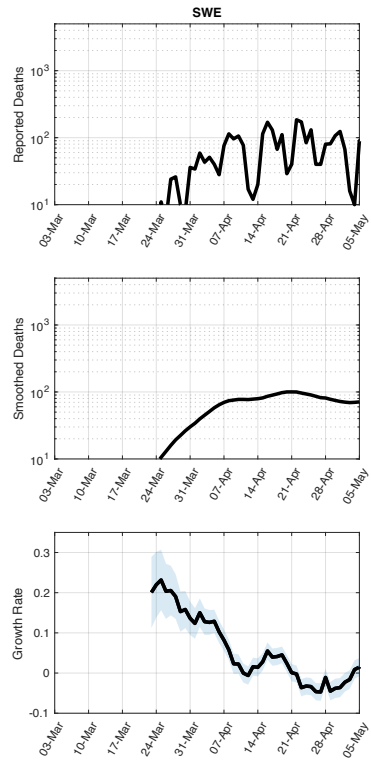

**Figure S3.** Plot of regression models fitted on 1,000 bootstrap replicates of the original data corresponding to the CovTM2 model.

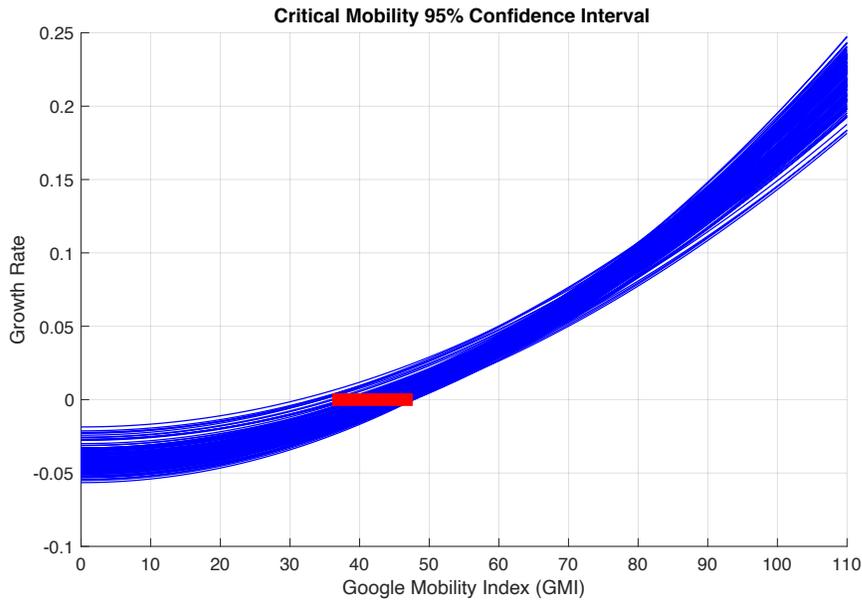

**Figure S4.** $R^2$ accuracy of regression models (order 2, without linear term) as function of delay Δ (measured in days) for 6 different categories of GMI.

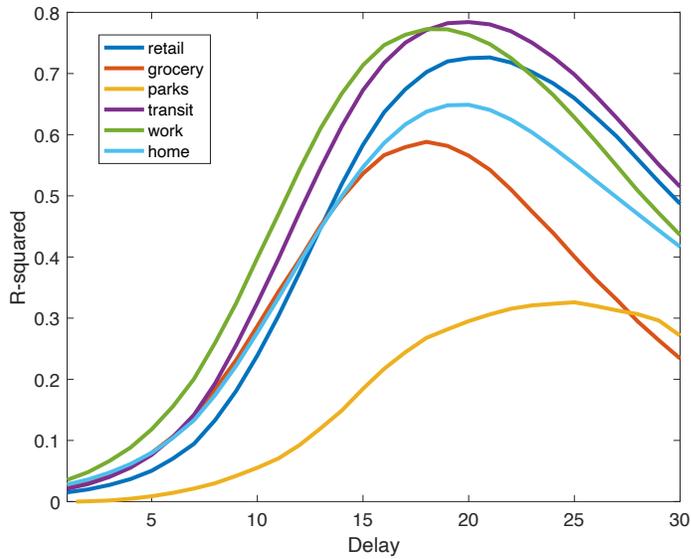

**Figure S5.** $R^2$ accuracy of regression models (order 2, without linear term) as function of delay Δ (measured in days) for 4 different data removal thresholds. The legend shows the threshold values, as well as (in parentheses) the number of remaining data points available for regression model fitting when the delay is 20 days.

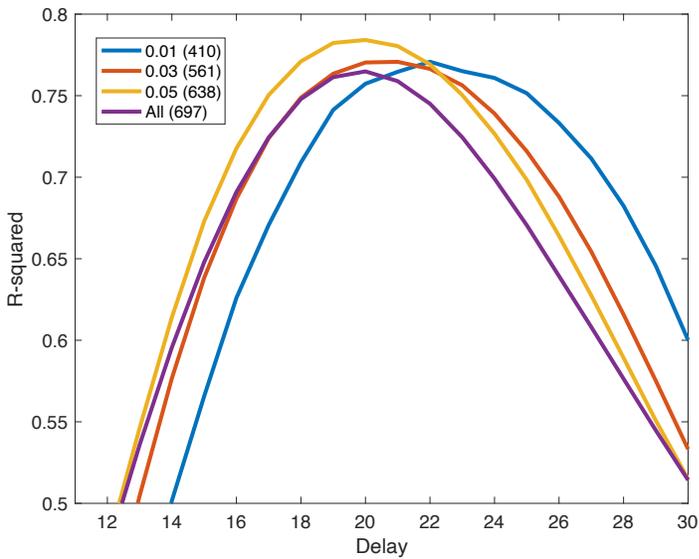

**Figure S6.** Polynomials of different degrees fitted to the data from 14 countries for a delay of 20 days, GMI transit mobility, and $\theta = 0.05$. Each circle is a data point (there are 636 data points after removal). The size of the circles corresponds to the weights of data points used for Weighted Least Squares (WLS).

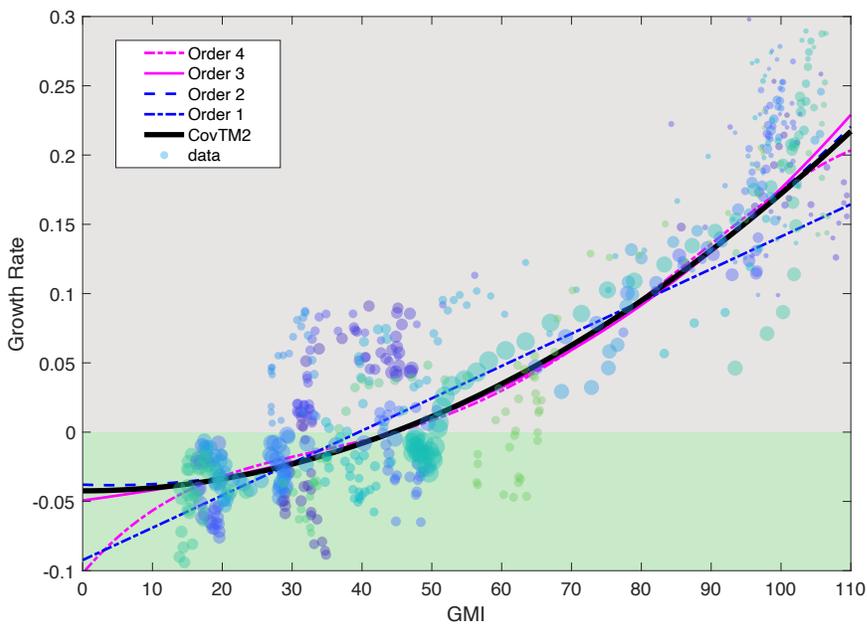

**Figure S7.** Country-specific regression models (14 figures for 14 countries). The description is the same as for Figure 3.

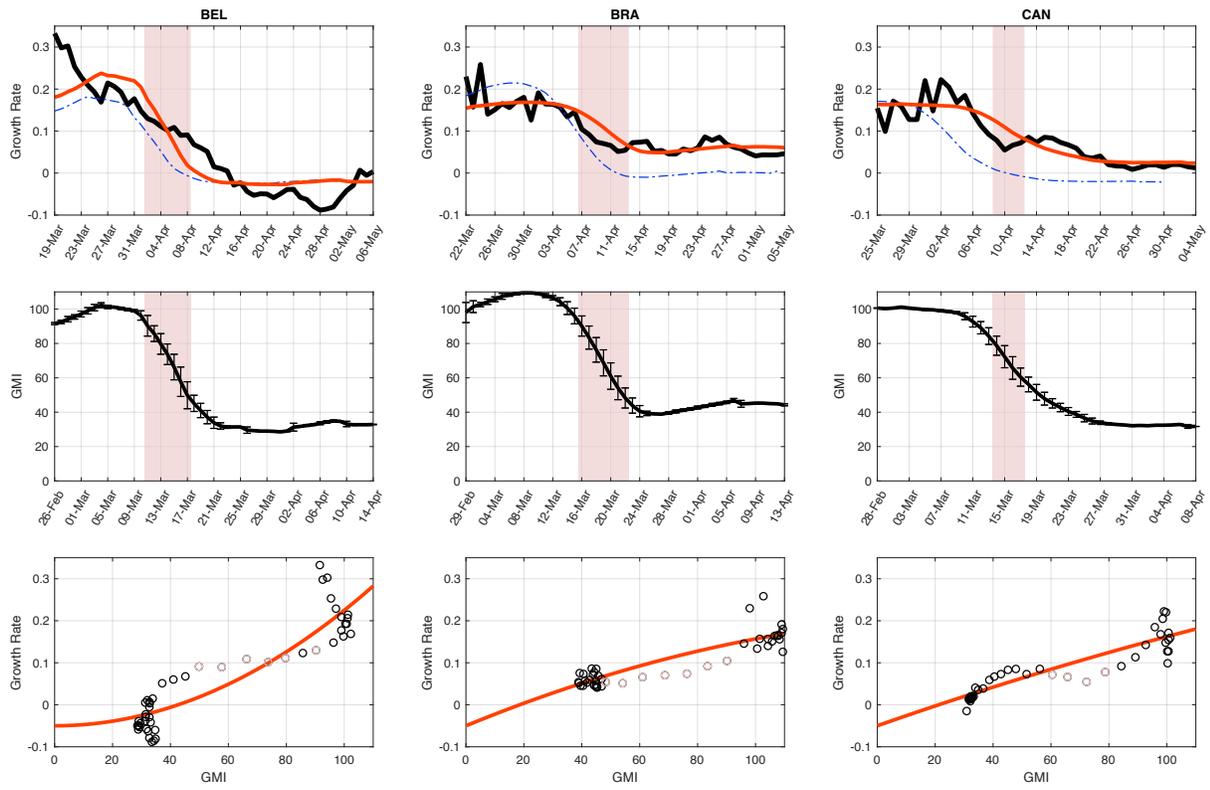

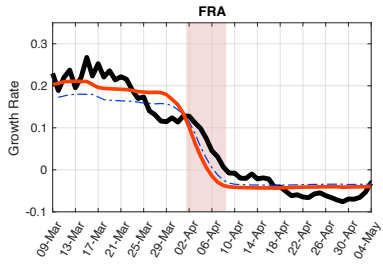
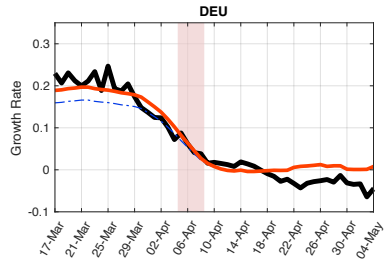
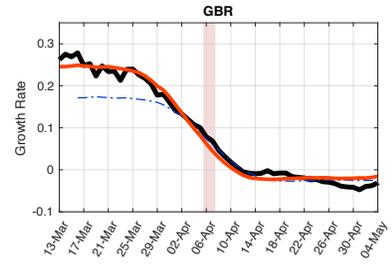
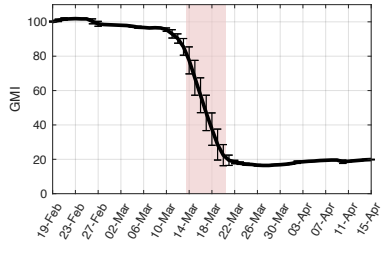
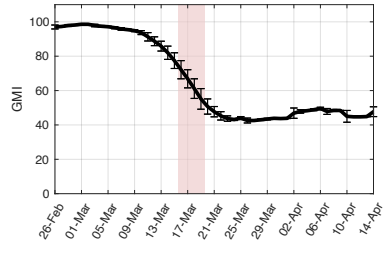
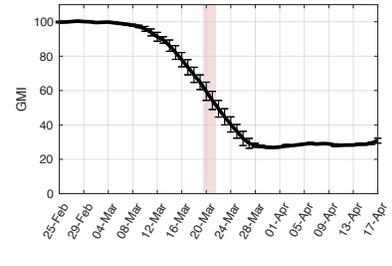
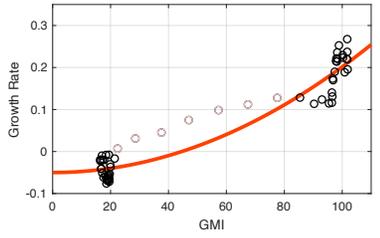
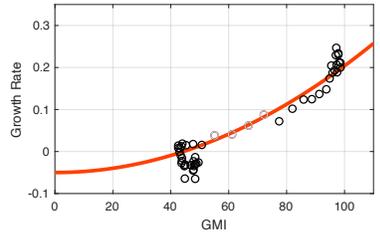
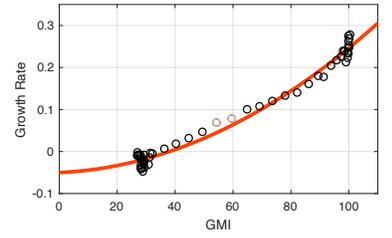
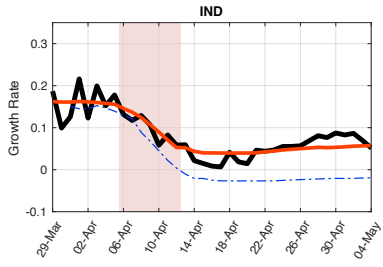
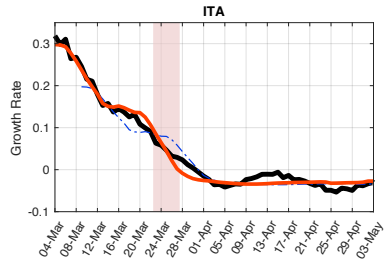
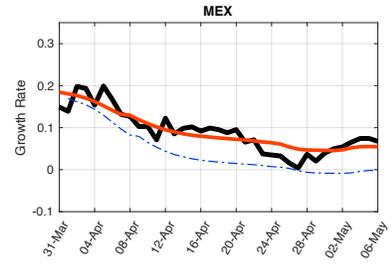
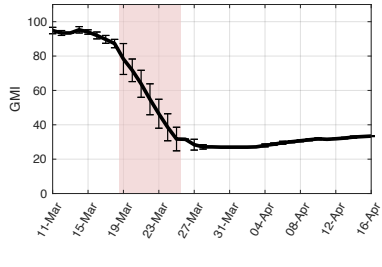
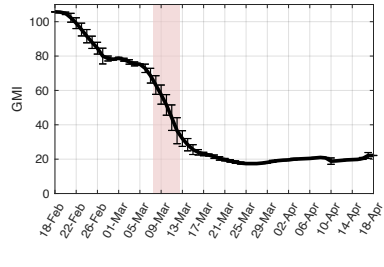
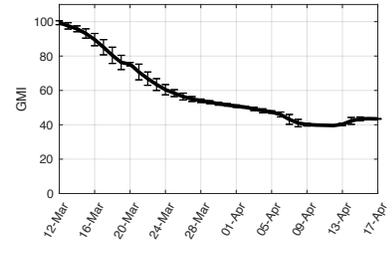
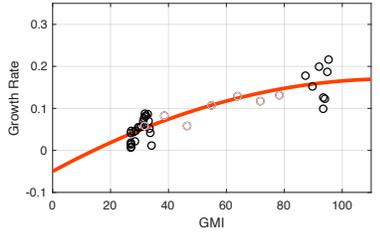
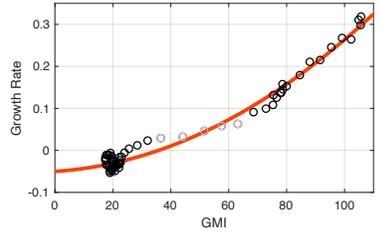
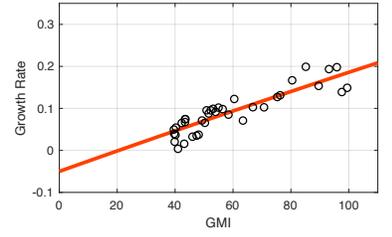

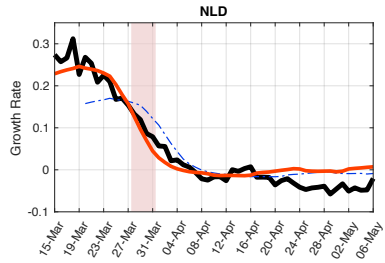
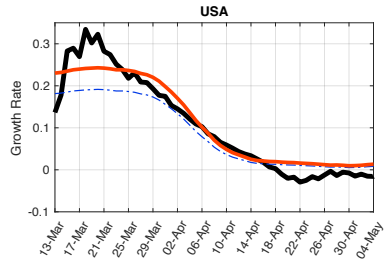
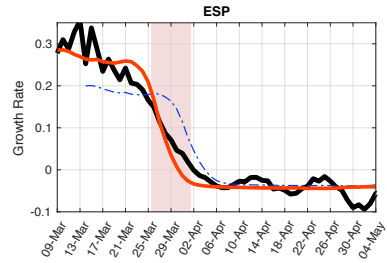
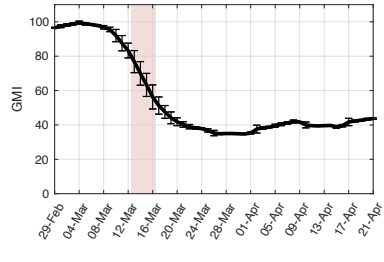
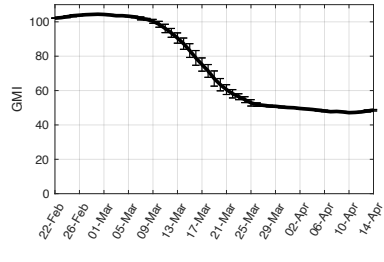
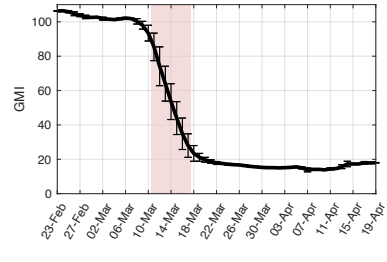
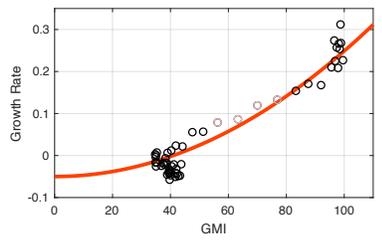
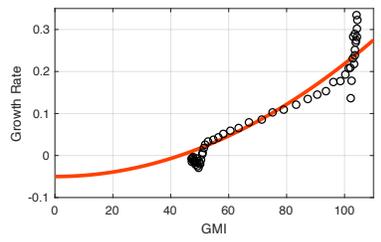
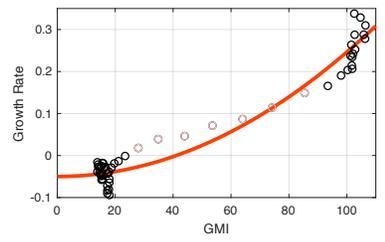
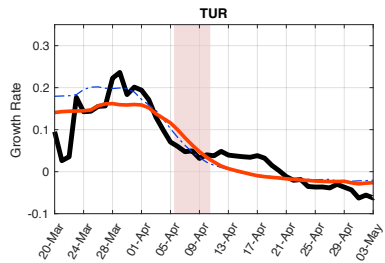
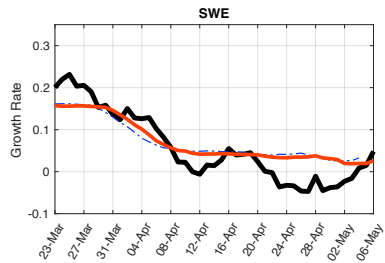
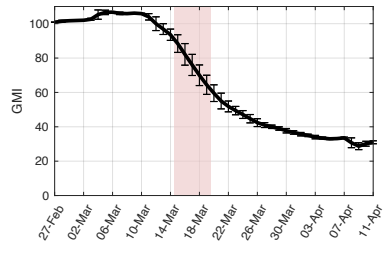
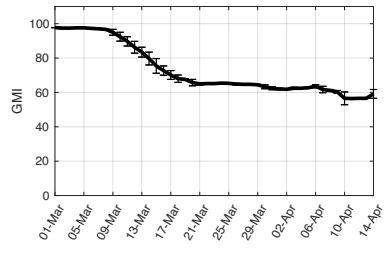
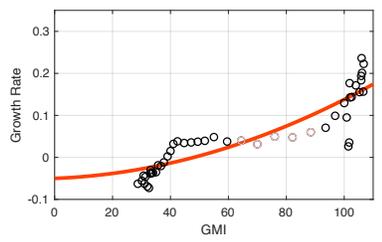
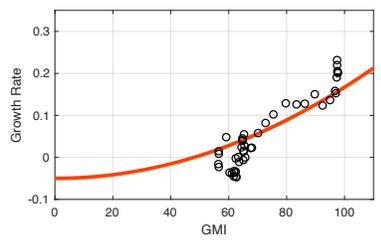

**Figure S8**. CovTM2d Regression Model: Relationship between COVID-19 infection growth rate and population mobility in 14 countries. Same explanation as for Figure 2.

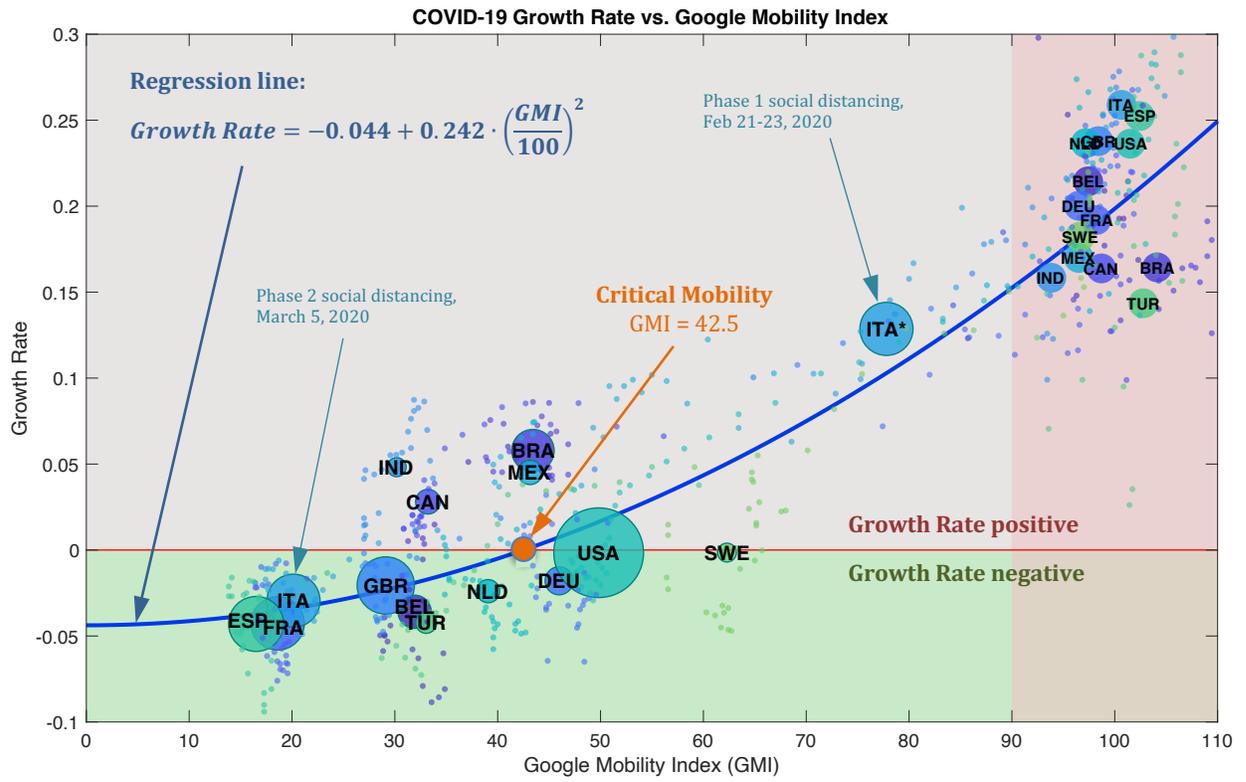